\documentclass[letterpaper, 10 pt, conference]{ieeeconf}  

\IEEEoverridecommandlockouts                              
\usepackage[utf8]{inputenc}
\usepackage[T1]{fontenc}

\usepackage{amsmath} 
\usepackage{amssymb}  

\usepackage{enumerate}
\usepackage{graphicx}
\usepackage{multirow}
\usepackage{array}
\usepackage{tikz}
\usetikzlibrary{quantikz2}
\usepackage[linesnumbered,ruled,vlined]{algorithm2e} 

\usepackage[colorlinks=true]{hyperref}
\usepackage{cleveref}
\usepackage[dvipsnames]{xcolor}

\title{\LARGE \bf
General circuit compilation protocol into partially fault-tolerant quantum computing architecture
}

\author{Tomochika Kurita$^{1}$
\thanks{$^{1}$Fujitsu Research of America, 4655 Great America Parkway Suite 410, Santa Clara, CA, United States, 95054
        {\tt\small kurita.tomo at fujitsu.com}}%
}

\begin{document}

\maketitle
\thispagestyle{empty}
\pagestyle{empty}

\begin{abstract}

As we are entering an early-FTQC era, circuit execution protocols with logical qubits and certain error-correcting codes are being discussed. Here, we propose a circuit execution protocol for the space-time efficient analog rotation (STAR) architecture. Gate operations within the STAR architecture is based on lattice surgery with surface codes, but it allows direct execution of continuous gates $Rz(\theta)$ as non-Clifford gates instead of $T = Rz(\pi/4)$. $Rz(\theta)$ operations involve creation of resource states $|m_\theta \rangle = \frac{1}{\sqrt{2}} (|0 \rangle + e^{i\theta} |1\rangle ) $ followed by ZZ joint measurements with target logical qubits. While employing $Rz(\theta)$ enables more efficient circuit execution, both their creations and joint measurements are probabilistic processes and adopt repeat-until-success (RUS) protocols which are likely to result in considerable time overhead. Our circuit execution protocol aims to reduce such time overhead by parallel trials of resource state creations and more frequent trials of joint measurements. By employing quadratic unconstrained binary optimization (QUBO) in determining resource state allocations within the space, we successfully make our protocol efficient. Furthermore, we proposed performance estimators given the target circuit and qubit topology. It successfully predicts the time performance within less time than actual simulations do, and helps find the optimal qubit topology to run the target circuits efficiently.

\end{abstract}

\section{Introduction}
Quantum computing is regarded as promising in the sense of solving such large-scale problems that classical computing cannot handle. Quantum gate operations, however, are affected by hardware noise, which limits the ability of quantum operation. Quantum error correction is the technique which corrects operation errors induced by hardware noise. This requires number of physical data qubits and measurement qubits to create one error-free logical qubits. There are a bunch of error-correcting codes to construct logical qubits, each of which has constraint for physical qubit topology, code distance, and transversal quantum operations.

Surface codes \cite{Fowler_2012} are nowadays the most considered error-correcting codes, where stabilizers can be locally located which makes them compatible with planar-type physical qubit topology like superconducting hardware. While we can only execute Pauli operations as transversal ones, we can execute all of universal quantum operations (H, S, T, CNOT) with lattice surgery \cite{Horsman_2012,Litinski_2018,Litinski_2019_Game}. Each of them is done with joint measurement of logical data qubits and ancilla qubits. Such a qubit-operating system, however, has several drawbacks. First, creating magic states which are required to execute T gates require many logical ancilla qubits (which is often called magic state factory) \cite{MSD_2012_Bravyi,Litinski_2019_MSD}. Second, some quantum operations can require many logical gates due to discreteness of basis gate sets \cite{Ross_2016_approx}.

To overcome such drawbacks, space time analog rotation (STAR) architecture \cite{STAR_ay1, STAR_ay2, STAR_tr1} is proposed, which is a modified lattice surgery system which replaces a $\mathrm{T} = \mathrm{Rz}(\pi/4)$ gate as the non-Clifford basis gate with an arbitrary-angle Z-axis analog rotation ($Rz(\theta)$). Such continuousness of the basis gate can significantly reduce the overall gates/steps to run target circuits. Also, the references \cite{STAR_ay1, STAR_tr1} introduced the resource states ($|m_\theta \rangle$) which are required to execute such analog rotations. This creation can be done within one apot for an single logical qubit.

There are two drawbacks of the STAR architecture. First, while the creating process of resource states involves considerable number of error-detecting steps, it is not fully fault-tolerant. The references \cite{STAR_ay1,  STAR_tr1} tackled this problem, and they proposed the carefully-designed protocols of creating resource states to maximally suppress the noise strengths. Second, such analog rotations involve probabilistic steps: (i) post-selection protocols in making high-fidelity resource states, (ii) analog rotations induced by joint measurements with resource states. The STAR architecture adopts a repeat-until-success (RUS) protocol for both steps, which can result in timestep overhead. Therefore, it is demanded to reduce/suppress such overhead.


There are numbers of papers regarding circuit execution protocol under the conventional surface code and lattice surgery \cite{Beverland_2022, Molavi_2025, Watkins_2024, Saravanan_2024}. However, there is a major difference in the circuit execution protocols. The conventional lattice surgery protocol assumes that the magic states $|m\rangle$ are generated in the magic state factory, which requires considerable number of ancilla logical qubits and is located at the side of logical qubit topology for main circuit executions. The STAR architecture, on the other hand, considers the creation process of resource states which is done within a spot for a single logical qubit. Therefore, we will need operation scheduling protocol specialized for the STAR architecture.


Note that the situation where the STAR architecture is useful is that the number of available physical qubits are still limited for running fully fault-tolerant quantum computation. In that sense, we would like to put as many logical qubits as possible into the physical qubit topology.

Here are the contributions this paper have:
\begin{itemize}
  \item{Devised operation scheduling protocol to run entire quantum circuits. In doing so, we carefully design how to do quantum operations: generating logical resource states (as ancilla states), moving ancilla states, joint measurement between data qubits and ancilla qubits, and two-qubit gates.}
  \item{Proposed performance estimators with regard to execution time, given some features of the target circuit and logical qubit topology. This helps take much less time to estimate the performance than actual simulations do, and find an optimal qubit topology to run the circuits efficiently.}
\end{itemize}

The structure of the rest of the paper is as follows. \Cref{sec:backgound} describes the backgrounds, \Cref{sec:op-protocols} describes the operation protocols we propose, \Cref{sec:res} describes the results we obtain, and \Cref{sec:conclusion} concludes the paper.

\section{\label{sec:backgound}Backgrounds}

In this section, we explain the details of quantum operations with the STAR architecture.
\subsection{Lattice surgery with Surface Code}
\begin{figure*}[htbp]
\centering
\includegraphics[scale=0.35]{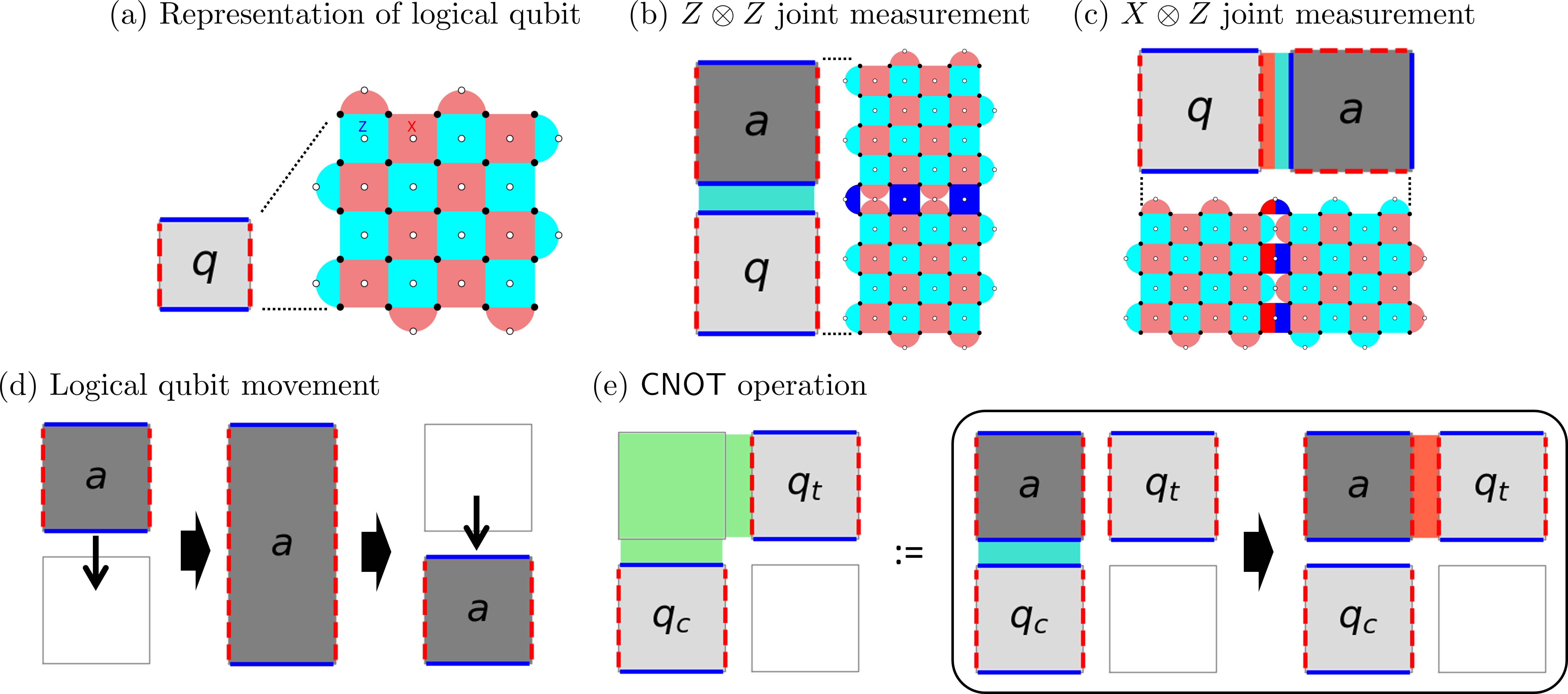}
\caption{\label{fig1} Description of quantum operation in surface code. (a) Description of a logical qubit in surface code. The code distance is $d=5$ in this figure. A logical qubit is described as one qubit "patch". Here, blue solid (red dashed) edges of the patch denotes the physical qubit array for Pauli-Z (Pauli-X) logical operations. (b,c) Schematic description of a $Z\otimes Z$ ($X\otimes Z$) joint measurement of a logical qubit $q$ and an ancilla qubit $a$. Such measurement is executed when the two qubits are adjacent across a Z-edge (an X-edge) of the data qubit and a Z-edge of the ancilla qubit, and by measuring the joint Pauli terms of Z (X) for the edge physical qubits of the logical data qubits and Z for the ones of the logical ancilla qubits. (d) Movement of logical qubits can be done by logical qubit expansion followed by its shrinkage. (e) A $\mathsf{CNOT}$ between two logical data qubits $q_c, q_t$ can be done with a logical ancilla qubit $a$ with its state of $|+\rangle$. First, $Z \otimes Z$ joint measurement of $q_c$ and $a$ is executed, then $X \otimes X$ joint measurement of $q_c$ and $a$. }
\label{fig:patches}
\end{figure*}

The surface code is an error-correcting code that possesses a high degree of compatibility with superconducting quantum computers. \Cref{fig:patches}a shows an example description of a logical qubit. It consists of $d^2$ physical data qubits (described with filled black circle) and $d^2 - 1$ physical measurement qubits (described with empty black circle), where $d$ denotes a code distance. A stabilizer is acting on physical data qubits located square-like (with their number of 4 at maximum), and a stabilizer check of $\mathrm{X}^{\otimes 4}$ ($\mathrm{Z}^{\otimes 4}$) is done by using an ancilla physical qubit located at the center of the red (blue) square. 

While the logical gate set which can be done transversally is limited to Pauli operations, other kinds of Clifford operations and non-Clifford operations can also be done with appropriate ancilla logical qubits via the lattice surgery setting.

Note that, when we execute operations with lattice surgery, we need an operation to stabilize the logical qubits after every single lattice-surgery operation. The stabilizing process involves $d$-time repetition of quantum error correction (QEC) cycles for all stabilizers. However, for some operations, such as qubit initialization to $|0\rangle$ and $|+\rangle$ as well as Pauli-flip operations, the state-stabilizing operations can be skipped since we can jointly correct errors occurred in such operations at the stabilizing operations just after the next operations. As the number of times of such stabilizing operation is the leading factor of the duration for circuit execution, we call such stabilizing operation a clock cycle, and we count the total number of clock cycles as the indicators of time performance of a whole circuit execution.

\Cref{fig:patches}b-e  shows the basic operations with lattice surgery. With this, each of the universal gate set {H,S,CNOT,T} is done in the following ways:

\nopagebreak

\begin{itemize}
  \item{An S gate, which is equivalent to Z-axis rotation by $\pi/2$, can be done by $Z\otimes Z$ joint measurement of the data logical qubits and $|+\rangle$ as the ancilla logical qubits (\Cref{fig1}b). The outcome of the measurement have to be +1 to succeed such operations, but even if it becomes -1, we can fix it by further applying a Pauli-Z gate to the target logical qubit. As $d$ cycles of QEC is required after $Z\otimes Z$ joint measurement, this operation takes 1 clock cycles.}
  \item{A CNOT gate is done with a logical qubit pair and a logical ancilla state in between (\Cref{fig1}e). First, $Z\otimes Z$ joint measurement between the control logical qubit and ancilla logical qubits is done, and then $X\otimes X$ joint measurement between the target logical qubit and the ancilla state. Since each joint measurement is followed by $d$ cycles of QEC, this operation takes 2 clock cycles.}
  \item{A Hadamard gate is done by two-time expansion and shrinkage of the logical qubits (\Cref{fig1}d) with the one-time movement of the edges (Not shown in \Cref{fig1}). This operation takes 3 clock cycles, since $d$ cycles of QEC is required after each of qubit expansions and edge movements.}
  \item{A T gate, which is equivalent to $\mathrm{Rz}(\pi/4)$, can be done in the same way as an S gate except that we need a magic state $|m \rangle = (|0\rangle + e^{i \pi /4}|1\rangle ) / \sqrt{2}$ as an ancilla logical state. The outcome of $Z\otimes Z$ joint measurement have to be +1 to succeed such operations, but even if it becomes -1, we can fix it by further applying an S gate to the target logical qubit. Note that creating a magic state fault tolerantly needs a lot of ancilla qubits and takes a large number of timesteps. For example, a well-known protocol to create a single magic state requires 15 ancilla logical qubits \cite{Litinski_2019_Game}.}
\end{itemize}

\subsection{STAR architecture}
STAR architecture offers a modified version of lattice surgery towards the early-FTQC quantum computation, largely in terms of non-Clifford operations. For the convebtional surface code and lattice-surgery architecture, the whole operation of a T gate is infeasible in the following two points. First, creating the fault-tolerant magic state requires many ancilla logical qubits, which significantly limits the number of logical qubits in the given space. Second, as all gates are discrete ones, even execution of a certain single-qubit operation requires more than a hundred basis gates \cite{Dawson_2006_SK}.
The STAR architecture introduces continuous gates $\mathrm{Rz}(\theta)$ as non-Clifford basis gates. (We call them analong gates hereafter.)

\begin{figure}[htbp]
\centering
\includegraphics[scale=0.35]{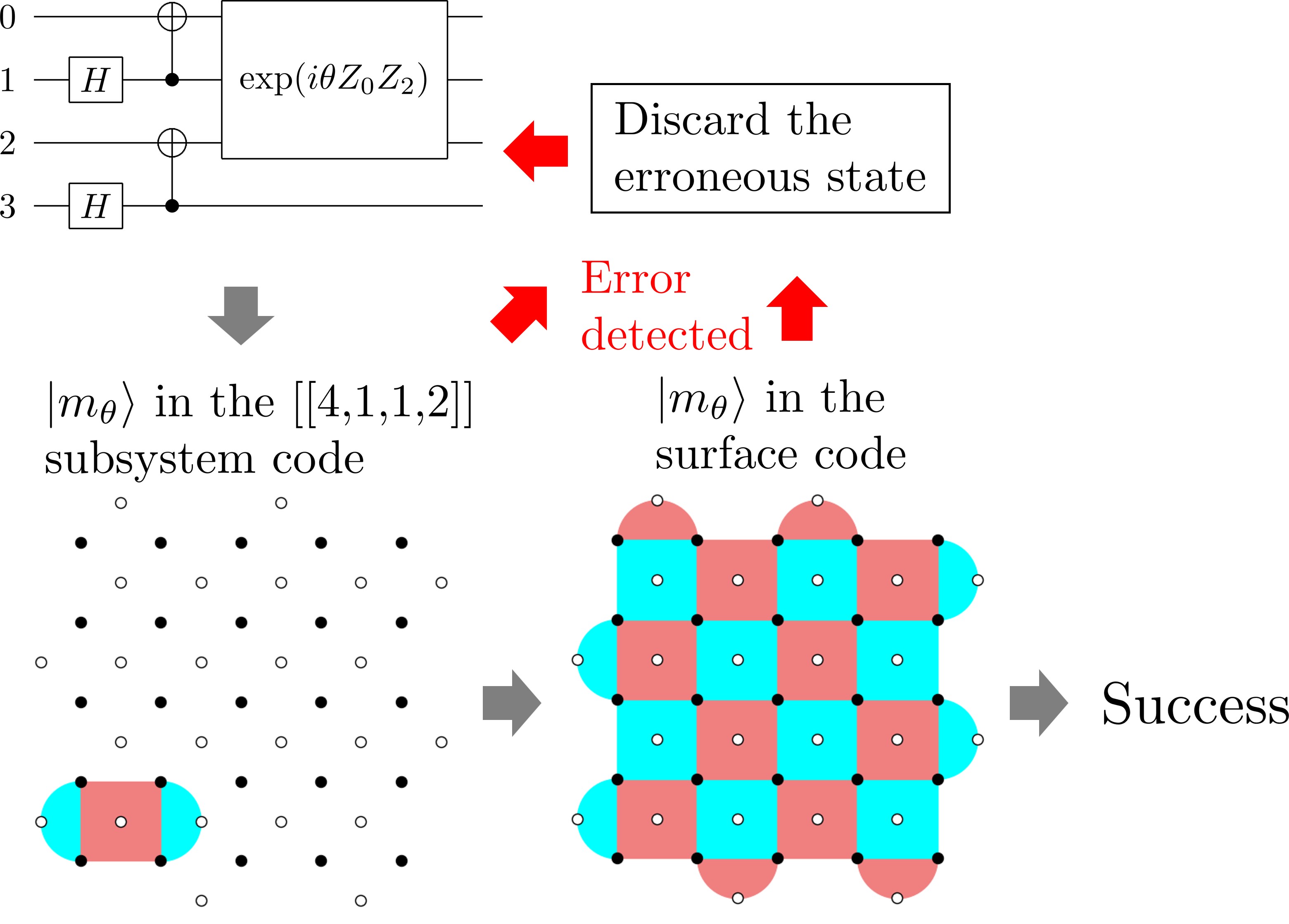}

\caption{\label{fig:rstatecr}Scheme of resource state ($| m_\theta \rangle$) creation, proposed in \cite{STAR_ay1}}
\end{figure}

\textbf{Creating resource states}. With the STAR architecture, such analog gates are executed via joint measurement of the target qubit and an ancilla state $| m_\theta \rangle = (|0\rangle + e^{i \theta}|1\rangle ) / \sqrt{2}$ which is called a resource state. Such a resource state is created within a spot for a single logical qubit. There are some proposed scheme to create resource state \cite{STAR_ay1,  STAR_tr1}, one of which is as follows \cite{STAR_ay1} (see \Cref{fig:rstatecr}):
\begin{enumerate}[(i)]
  \item A physical qubit with the state $|m_\theta \rangle$ is created.
  \item A logical qubit with $2\times2$ physical qubits is created employing [[4,1,1,2]] subsystem code.
  \item Expand the logical qubit with the desired code distance \cite{Horsman_2012}
\end{enumerate}

Syndrome measurements are executed in the steps (ii) and (iii) to detect errors. We discard the states found out to be erroneous in the steps (ii) and (iii), and repeat the resource state creation until one passes both syndrome measurements in (ii) and (iii). Due to such a post-selection process, the creation process is probabilistic. We define the success probability as $p_\mathrm{cr}$. Its value depends on the error rates of the physical qubits and the code distance we adopt. 

The whole process of creating resource states with syndrome measurements makes the resultant resource states partially robust, in the sense that the rates of logical errors are reduced to $P_Z = 2p/15 + O(p^2)$ and $P_X = O(p^2)$, where $p$ denotes the error rate of a physical qubit \cite{STAR_ay1}. With another scheme of resource state preparation \cite{STAR_tr1}, the rate of phase-flip errors is reduced to $O(p\theta)$, where $\theta$ denotes the angle of Z-rotation applied to make a resource state. We have to consider some error-mitigation strategies to reduce the effect of the remnant errors to obtain meaningful results, while such a field is out of scope for this work.

\begin{figure}[htbp]
\centering
\begin{quantikz}
\lstick{\ket{\psi}} & \gate[2]{M_{P\otimes\mathrm{Z}}} &\gate{P} \wire[d][1]{c}  & \rstick{$R_{P}(\pm \theta)$ \ket{\psi}}\\
\lstick{\ket{m_{\theta}}} &       & \gate{M_\mathrm{X}} 
\end{quantikz}

\caption{\label{qcir:analogrotation}Scheme of analog rotation $R_P(\theta)$ with a resource state $| m_\theta \rangle$, where $P$ denotes the Pauli basis of the analog rotation. The rotation $R_P(+\theta)$ becomes successful if the measurement outcome of $M_{P \otimes {\mathrm{Z}}}$ is $+1$, and $R_P(-\theta)$ acts if the outcome is $-1$.}
\end{figure}
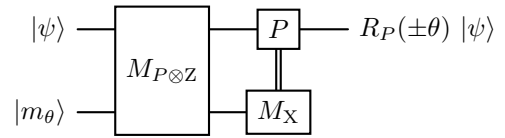

\textbf{Analog rotations with resource states}. Analog rotations $\mathrm{R}_P(\theta)$ are executed through $(P\otimes \mathrm{Z})$-basis joint measurements of the target logical state  $|\psi \rangle$ and the ancilla state $|m_\theta \rangle$, as shown in \Cref{qcir:analogrotation}. Here, $P$ denotes a Pauli term acting on the target qubit(s). It is also a probabilistic process in a sense that such rotation are successful only if the outcome of the joint measurement is +1. If it results in $-1$, the operation results in $\mathrm{R}_P(-\theta)$. In the latter case, we have to further perform $\mathrm{R}_P(2\theta)$, whose angle is a sum of $+\theta$ as compensation of preceding $-\theta$, and additional $+\theta$ as the final rotation angle we desire. More generally, for $m$th trial of analog rotations, we have to try $\mathrm{R}_P(2^{m-1}\theta)$ with the resource state $|m_{2^{m-1}\theta }\rangle$. We have to repeatedly try such a analog rotation until it results in success. The success probability of a single trial is $p_{\mathrm{cm}}=0.5$.

\section{\label{sec:op-protocols}Details of operation protocols} 
In this section, we explain the basis gate set we employ.
Here we set some guidelines regarding this, as follows:
\begin{enumerate}[(a)]
\item Each analog rotation gate acts on a single qubit. 
\item Refrain from using Hadamard gates.
\end{enumerate}

For the guideline (a), we may want to execute as many analog rotations as possible in parallel, to maximally reduce the performance overhead. In this sense, single-qubit operations are most preferred. While simultaneous rotations can be done even when we allow multi-qubit analog rotation gates, there are some limitations: (i) any two of simultaneous gates cannot share qubits to act on, and (ii) any two of simultaneous gates cannot block  the qubit routing of each other.
For the guideline (b), Hadamard gates are costly, in the sense that it takes up to 3 clock cycles and involves logical qubit reformation.

Based on such consideration, we set the basis gate set to be ${\mathsf{CNOT}, \mathrm{Rz}(\theta), \mathrm{Rx}(\phi)}$. We use the fact that any kinds of single-qubit gates can be decomposed with three arbitrary-angle rotation gates around X-, Y- or Z-axes. In this paper, we use the circuit transpilation where any single-qubit operations is described as $Rz(\theta)Rx(\phi)Rz(\psi)$. Here, Rz(Rx) is done by ZZ(XZ) joint measurement of a target state and resource state.

Another thing to consider is how to create a rule for each operation of lattice surgery to reserve empty spots for logical qubits. Here is the list of operations which need empty spots:
\begin{enumerate}[(1)]
  \item Movement of resource states to adjacent spots of the corresponding data qubits
  \item CNOT routing between logical data qubits
  \item  Creation trials of resource states
\end{enumerate}
While we can dynamically change the priority of these element operations, in this paper, we set the highest priority on the operation (1), then (2) as the second highest and (3) as the lowest. Our observation shows that we can execute any kind of quantum circuits without failure with this priority order.

This order is determined with a criteria that how long (i.e. how many clock cycles) each operation take and how likely the operation can be blocked by other operations. Regarding operations (1), it takes only one clock cycle. In addition, as we limit the spots for resource state creation to distance 2 or lower from the target logical qubit, it is not likely it blocks other operations. Regarding operations (2), it takes 2 clock cycles and the path can be long especially if the corresponding two logical qubits are located far away from each other. Therefore, the path can be easily blocked by other kinds of operations. And it can be the case that we cannot make paths between the logical qubits especially if the logical qubits are densely mapped into the logical qubit topology. 

Here we break down the operation protocols. We design the operations for every timestep $t$. First, we categorize the data logical qubits into one of the categories shown in \Cref{tab:qcat}. We denote the set of qubits in one category with  $\mathcal{Q}_{\{ \mathtt{category\_name} \}}$, such as $\mathcal{Q}_{\mathtt{IDLE}}$, $\mathcal{Q}_{\mathtt{CNOT1}}$ and so on.

\begin{table*}[htbp]
\centering
\caption{\label{tab1} Categorization of quantum operations with lattice surgery and STAR architecture} 
\label{tab:basisop}
\begin{tabular}{|p{4.5cm}|p{12cm}|}
\hline
Category & Explanation 
\\
\hline
Qubit initialization with Pauli-eigenstate & 0 clock cycle, since errors can be jointly corrected with one occured in the subsequent operation. \\
\hline
Qubit movement & 1 clock cycle, where stabilization is required between qubit expansion and shrinkage \\
\hline
Resource state creation & 1 clock cycle with success probability of $p_{\mathrm{cr}}$. Note that we can regard this process less than 1 clock cycle when the creation trial results in failure. When $\theta = \pm \pi /2$, which is an eigenstate of Pauli, it is regarded as 0 cycle and always successful ($p_{\mathrm{cr}}=1$)\\
\hline
Analog rotation with resource state & 1 clock cycle with success probability of $p_{\mathrm{cm}} = 0.5$\\
\hline
CNOT & 2 clock cycles. First stabilization occurs after the $\mathrm{Z} \otimes \mathrm{Z}$ joint measurement, and second one occurs after the $\mathrm{X} \otimes \mathrm{X}$ joint measurement\\
\hline
\end{tabular}
\end{table*}

\begin{table}[htbp]
\centering
\caption{\label{tab1} Qubit categorization} 
\label{tab:qcat}
\begin{tabular}{|wc{1.2cm}|p{6cm}|}
\hline
Category & Explanation 
\\
\hline
$\mathtt{IDLE}$ & All operations acting on this qubit are already executed, or waiting for another qubits to execute the following CNOT operation  \\
\hline
$\mathtt{CNOT1}$ & The first timestep of CNOT operation  \\
\hline
$\mathtt{CNOT2\_RX}$ & The second timestep of CNOT operation, and RX operation follows  \\
\hline
$\mathtt{CNOT2\_RZ}$ & The second timestep of CNOT operation, and RZ operation follows  \\
\hline
$\mathtt{CNOT2}$ & The second timestep of CNOT operation, and none of RX operation and RZ operation follows \\
\hline
$\mathtt{RX1}$ & In the process of RX operation, and the corresponding resource state is not ready  \\
\hline
$\mathtt{RX2}$ & In the process of RX operation, and the corresponding resource state is ready but not located at the adjacent place  \\
\hline
$\mathtt{RX3}$ & In the process of RX operation, and the corresponding resource state is ready and located at the adjacent place  \\
\hline
$\mathtt{RZ1}$ & In the process of RZ operation, and the corresponding resource state is not ready  \\
\hline
$\mathtt{RZ2}$ & In the process of RZ operation, and the corresponding resource state is ready but not located at the adjacent spot  \\
\hline
$\mathtt{RZ3}$ & In the process of RZ operation, and the corresponding resource state is ready and located at the adjacent spot  \\
\hline
\end{tabular}

\end{table}

\begin{algorithm}[htbp]
\SetAlgoLined
\KwIn{Set of all logical qubits $\mathcal{Q}$; Set of all spots for qubits $\mathcal{S}$, Set of logical data qubits and their positions $\{ (q_i$, $\vec{v}_i ) \}$ where $\vec{v}_i := (x_i,y_i)$; Set of the spots which are already in use $\{ \vec{u} \} $}
\KwOut{Spots for resource-state creation trial $\{\vec{r}\}_{q_i}$ for each $q_i \in \mathcal{Q}_\mathtt{RX1} \cup \mathcal{Q}_\mathtt{RZ1}$}
$\mathcal{S} \leftarrow \mathcal{S} \backslash (\{ \vec{v}_i\} \cup \{\vec{u}\})$\;

\For{$q_i \in \mathcal{Q}_\mathtt{RX2} \cup \mathcal{Q}_\mathtt{RZ2}$}{
  Move the corresponding resource state to the position adjacent to $q_i$ \;
  $\{ \vec{w} \} \leftarrow $ Set of all positions used in this movement\;
  $\mathcal{S} \leftarrow \mathcal{S} \backslash \{ \vec{w}\}$
}
\For{$(q_c,q_t)\in \mathcal{Q}_{\mathtt{CNOT1}}$ }{
  Try CNOT routing from $q_c$ to $q_t$ \;
  \If{Routing is successful}{
    $\{ \vec{w} \} \leftarrow $ Set of all positions for qubits used in this routing\;
    $\mathcal{S} \leftarrow \mathcal{S} \backslash \{ \vec{w}\}$
  }
}
$\mathbf{b} = \mathrm{CreateBinSet\_1st}(\mathcal{Q}, \mathcal{S})$\;
$f = \mathrm{CreateBinEquation\_1st}(\mathbf{b})$\;
Determine the values of $\mathbf{b}$ which minimize $f$\;
$\mathbf{b}_{\mathrm{eff}} = \{b_j \in \mathbf{b}| b_j=1 \}$ \;
$\mathbf{c} = \mathrm{CreateBinSet\_2nd}(\mathcal{Q}, \mathcal{S},\mathbf{b}_{\mathrm{eff}})$\;
$g = \mathrm{CreateBinEquation\_2nd}(\mathbf{c})$\;
Determine the values of $\mathbf{c}$ which minimize $g$\;
$\mathbf{c}_{\mathrm{eff}} = \{c_j \in \mathbf{c}| c_j=1 \}$ \;
$\mathbf{b}_{\mathrm{eff}} \leftarrow \mathbf{b}_{\mathrm{eff}} \cup \mathbf{c}_{\mathrm{eff}}$\;
\For{$q_i \in \mathcal{Q}_\mathtt{RX1} \cup \mathcal{Q}_\mathtt{RZ1} \cup \mathcal{Q}_\mathtt{RX3} \cup \mathcal{Q}_\mathtt{RZ3} \cup \mathcal{Q}_\mathtt{CNOT\_RX} \cup \mathcal{Q}_\mathtt{CNOT\_RZ}$}{
  Let $\mathbf{b}_{\mathrm{eff},q_i} \subset \mathbf{b}_{\mathrm{eff}}$ be a set of binary variables corresponding to qubit $q_i$\;
  Let $\{\vec{r}\}_{q_i}$ be a set of positions corresponding to $\mathbf{b}_{\mathrm{eff},q_i}$\;
}

\caption{Overall algorithm}
\end{algorithm}

\begin{algorithm}[htbp]
\caption{CreateBinarySet\_1st}\label{alg:bin_cr1}
\SetAlgoLined
\KwIn{Set of logical qubits $\mathcal{Q}$; Set of positions for qubits $\mathcal{S}$, Set of logical data qubits and their positions $\{ (q_i$, $\vec{v}_i ) \}$ where $\vec{v}_i := (x_i,y_i)$}
\KwOut{Set of binary variables $\mathbf{b}$}
$\mathbf{b}\leftarrow \emptyset, j \leftarrow 0$\;
\For{$q_i\in {\mathcal{Q}_{\mathtt{RX1}}} \cup \mathcal{Q}_\mathtt{RX3} \cup \mathcal{Q}_\mathtt{CNOT\_RX} $ }{
  \For{$y\in \{ y_{i}-1, y_{i}+1\}$}{
    \If{$(x_i,y) \in \mathcal{S}$}{
    Set a binary variable $b_j\in\{0,1\}$ which denotes that we try to create resource states at the location $(x_i,y)$ for qubit $q_i$\;
    $\mathbf{b}\leftarrow \mathbf{b} + \{b_j\}$\;
    $j \leftarrow j+1$
    }
  }
}
\For{$q_i\in {\mathcal{Q}_{\mathtt{RZ1}}} \cup \mathcal{Q}_\mathtt{RZ3} \cup \mathcal{Q}_\mathtt{CNOT\_RZ}$ }{
  \For{$x\in \{ x_{i}-1, x_{i}+1\}$}{
    \If{$(x,y_i) \in \mathcal{S}$}{
    Set a binary variable $b_j\in\{0,1\}$ which denotes that we try to create resource states at the location $(x,y_i)$ for qubit $q_i$
    $\mathbf{b}\leftarrow \mathbf{b} + \{b_j\}$\;
    $j \leftarrow j+1$
    }
  }
}
\end{algorithm}

\begin{algorithm}[htbp]
\caption{CreateBinaryEquation\_1st}\label{alg:bin_eq1}
\SetAlgoLined
\KwIn{Set of all logical qubits $\mathcal{Q}$; Set of positions for qubits $\mathcal{S}$, Set of binary variables $\boldsymbol{b}$}
\KwOut{Binary equation $f$ to solve}
$ f_1 \leftarrow - \sum_{b\in\mathbf{b}} b$ \;
$f_2 \leftarrow 0,f_3 \leftarrow 0$\;
\For{$\vec{v} \in \mathcal{S}$}{
  Let $\boldsymbol{b}_{\vec{v}} \subset \boldsymbol{b}$ which correspond to position $\vec{v}$ \;
  \If{$|\boldsymbol{b}| \ge 2$}{
    \For {$b_k,b_l \in \boldsymbol{b}, k<l$}{
    $f_2 \leftarrow f_2 + b_k b_l$
    }
  }
}
\For{$q_i \in \mathcal{Q}$}{
  Let $\boldsymbol{b}_{q_i}:=\{b_{a_0}\cdots b_{a_k}\} \subset \boldsymbol{b}$ which correnspond to qubit $q_i$\;
  $f_3 \leftarrow f_3 + \prod_{l=0}^{k} (1-b_{a_l})$
}
$f \leftarrow f_1 + f_2 +f_3$

\end{algorithm}

\begin{algorithm}[htbp]
\SetAlgoLined
\KwIn{Set of logical qubits $\mathcal{Q}_{\mathtt{RX1}},\mathcal{Q}_{\mathtt{RZ1}}$; Set of positions for qubits $\mathcal{S}$; $\mathbf{b}_{\mathrm{eff}}$}
\KwOut{Set of binary variables $\mathbf{c}$}
$\mathbf{c}\leftarrow \emptyset, j \leftarrow 0$\;

\For{$q_i\in {\mathcal{Q}_{\mathtt{RX1}}} \cup {\mathcal{Q}_{\mathtt{RZ1}}}$ }{
  Let $\boldsymbol{b}_{\mathrm{eff},q_i} \subset \boldsymbol{b}_{\mathrm{eff}}$ be a set of binary variables corresponding to qubit $q_i$\;
  Let $\{ \vec{w}_1\}$ be a set of positions corresponding to $\boldsymbol{b}_{\mathrm{eff},q_i}$\;
  \For{$\{\vec{w}_2 | \vec{w}_2 \in \mathcal{S}, \| \vec{w}_2-\vec{w}_1 \|_1 = 1\}$}{
    Set a binary variable $c_j\in\{0,1\}$ which denotes that we try to create resource states at the location $\vec{w}_2$ for qubit $q_i$\;
    $\mathbf{c}\leftarrow \mathbf{c} + \{c_j\}$\;
    $j \leftarrow j+1$
  }
}
\caption{CreateBinarySet\_2nd}
\end{algorithm}

\begin{algorithm}[htbp]

\caption{CreateBinaryEquation\_2nd}\label{alg:bin_eq2}
\SetAlgoLined
\KwIn{Set of all logical qubits $\mathcal{Q}$; Set of positions for qubits $\mathcal{S}$, Set of binary variables $\mathbf{c}$}
\KwOut{Binary equation $g$ to solve}
$ g_1 \leftarrow - \sum_{c\in\mathbf{c}} c$ \;
$g_2 \leftarrow 0$\;
\For{$\vec{v} \in \mathcal{S}$}{
  Let $\mathbf{c}_{\vec{v}} \subset \mathbf{c}$ which correspond to position $\vec{v}$ \;
  \If{$|\mathbf{c}_{\vec{v}}| \ge 2$}{
    \For {$c_k,c_l \in \mathbf{b}, k<l$}{
    $f_2 \leftarrow g_2 + c_k c_l$
    }
  }
}

$g \leftarrow g_1 + g_2$

\end{algorithm}

\subsection{Basis operations}
Here we define how to manage the operation in the architecture. First, we describe the qubit spaces as a grid graph, as shown in \Cref{fig:qcoord}. Here, any spot is described as a 2D lattice point $\vec{v} = (m, n)$ where $m,n \in \mathbb{Z}_{\ge 0}$ and the location of the bottom left spot is $(0,0)$. We set a distance between adjacent spots $\| \vec{u}-\vec{v} \|_1$ to be 1.

In quantum operations with surface code and lattice surgery, the leading factor of operation time is the number of times we perform stabilization of logical qubits. Such stabilization generally takes $d$ cycles of QEC. We define the time for stabilizing a logical qubit as one clock cycle. We summarize the basis operations and how many clock cycles are required for each operation in \Cref{tab:basisop}. Based on this, we count the number of clock cycles as a performance on target circuits.

\begin{figure}[htbp]
\centering
\includegraphics[scale=0.4]{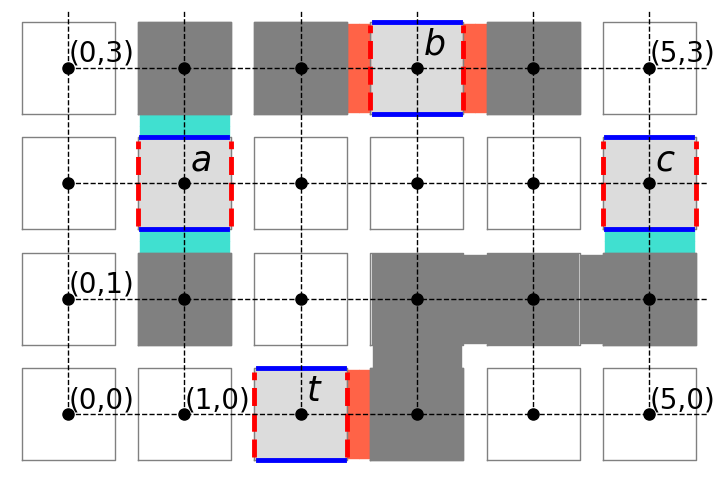}
\caption{\label{fig:qcoord}Grid graph representing qubit spaces. Here every logical qubit is put so that the X-edges are on the left and right sides and the Z edges are on the top and bottom sides. (a) For $\mathrm{Z}\otimes P$ joint measurement with a data qubit placed in $(m,n)$, an ancilla state is placed in $(m,n \pm 1)$. (b) Similarly, for $\mathrm{X}\otimes P$ joint measurement with a data qubit placed in $(m,n)$, an ancilla state is placed in $(m\pm 1,n )$. (c,t) To perform CNOT between two data qubits, we create an ancilla state of $|+\rangle$ which connects to a Z-edge of the control qubit via and its Z-edge, and to an X-edge of the target quibit via its X-edge. The spots denoted with darkgray are the candidate positions of ancilla qubits.
}
\end{figure}

\subsection{Spot allocation of resource creation trials}

We will need some algorithms to allocate spots for resource states. As their creation involves probabilistic procedure which can result in operation time overhead, we may want to do multiple creation trials in parallel. Therefore, we may need as many spots as possible per target resource state and in some cases we may encounter some conflicts. We have some ways to alleviate this: (i) allocate a logical data qubit as well as the rotation Pauli basis for each available spot, or (ii) dynamically optimize the location. The way (i) may get rid of the conflicts for spot usage, but it cannot be applied when there is an empty spot adjacent to two logical qubits or more. Therefore, we choose the way (ii). To take this way, we create an equation to minimize, with multiple binary variables: $\boldsymbol{b} = \{ b_i|b_i \in \{0,1\} \text{ for all }i\}$. A single variable is allocated per pair of a target qubit and a neighboring empty spot for creating a resource state. The equation is as follows:

\begin{align}
\label{eq:bin_1}
f(\boldsymbol{b}) = f_1+Af_2+Bf_3,
\end{align}

where $A,B\in \mathbb{R}_{+}$ denote some constants. Each terms in \Cref{eq:bin_1} is created in \Cref{alg:bin_eq1}, with underlying concepts as follows: 
\begin{enumerate}
  \item[$f_1$: ] Prepare as many resource states as possible
  \item[$f_2$: ] Cannot place two or more resource states in one spot. 
  \item[$f_3$: ] Prepare at least one resource state per data qubit, if possible
\end{enumerate}

The terms $f_2$ and $f_3$ are regarded as penalties, and indeed the value of these terms are zero when the underlying conditions of $f_2$ and $f_3$ are met. We can find the optimal value set $\boldsymbol{b}$ to minimize $f ( \boldsymbol{b} )$ by employing annealing optimization with classical computer \cite{Bertsimas_1993_SA}, quantum annealer \cite{dwave_2011} and quantum-inspired annealer \cite{Aramon_2019_DA}. Note that the maximum degree of variables is 2 in both terms, $f_2$ and $f_3$. 

Furthermore, we can try additional resource-state creation in a spot of distance 2 form the target logical qubit. After determining $\boldsymbol{b}$ which minimize $f(\boldsymbol{b})$, we create an equation to minimize with multiple binary variables $\boldsymbol{c} = \{ c_i|c_i \in \{0,1\} \text{ for all }i\}$. A single variable is allocated per pair if target qubit and a distance-2 spot for creating a resource state. The equation is as follows:
\begin{align}
\label{eq:bin_2}
g(\boldsymbol{c}) = g_1 + A g_2
\end{align}
Each of the terms is created in \Cref{alg:bin_eq2}, with underlying concepts as follows: 
\begin{enumerate}
  \item[$g_1$: ] Prepare as many resource states as possible
  \item[$g_2$: ] Cannot place two or more resource states in one spot. 
\end{enumerate}
We can also include a constraint like $g_3$:
\begin{enumerate}
  \item[$g_3$: ] Prepare at least one resource state per data qubit, if possible
\end{enumerate}
but do not include it in \Cref{eq:bin_2} and \Cref{alg:bin_eq2} since we are going to guarantee at least one spot for resource state creation in the preceding optimization of $f(\boldsymbol{b})$ in \Cref{eq:bin_1}. Furthermore, incorporating it can lead to greater degree of variables than 2, which makes some annealers unable to handle $g(\boldsymbol{c})$ as is. Note that, even in this case, we can employ degree-reduction protocols so that the modified equation has maximum degree of 2 by adding extra number of binary variables \cite{Verma_2021_QUBO}.

\subsection{Frequent attempts of analog rotation} 
As mentioned in the previous section, reducing the interval of joint measurement trials is another strategy of reducing time overhead. Specifically, in the case both of the two $P$-edges are accessible for joint measurement, where $P \in \{ \mathrm{X},\mathrm{Z} \}$, we can create a resource state $|m_{2 \theta} \rangle$ for the next measurement trial at one side while executing joint measurement of $|m_{\theta} \rangle$ and the target qubit at the other side, which makes the interval 1 clock cycle. Otherwise, the interval will take at least 2 clock cycles: one clock cycle for creating a resource state at the location or moving of a resource state from another location, and another clock cycle for joint measurement. In such a sense, we include $q_i \in \mathcal{Q}_{\mathtt{RX3}} \cup \mathcal{Q}_{\mathtt{RZ3}}$ in the creating process of binary variables in \Cref{alg:bin_cr1}, as well as $q_i \in \mathcal{Q}_{\mathtt{RX1}} \cup \mathcal{Q}_{\mathtt{RZ1}}$.

Similarly, when a qubit is in the process of CNOT operation and its next operation is an analog single-qubit rotation, we can reduce the intervals between the CNOT operation and the following joint measurement with a resource state, by running resource state creation trials in the second clock cycle of CNOT operation. By doing this, we can do joint measurement just after the CNOT operation is done. In such a sense, we also include $q_i \in \mathcal{Q}_{\mathtt{CNOT\_RX}} \cup \mathcal{Q}_{\mathtt{CNOT\_RZ}}$ in the creating process of binary variables in \Cref{alg:bin_cr1}.

\begin{figure}[htbp]
\centering
\includegraphics[scale=0.8]{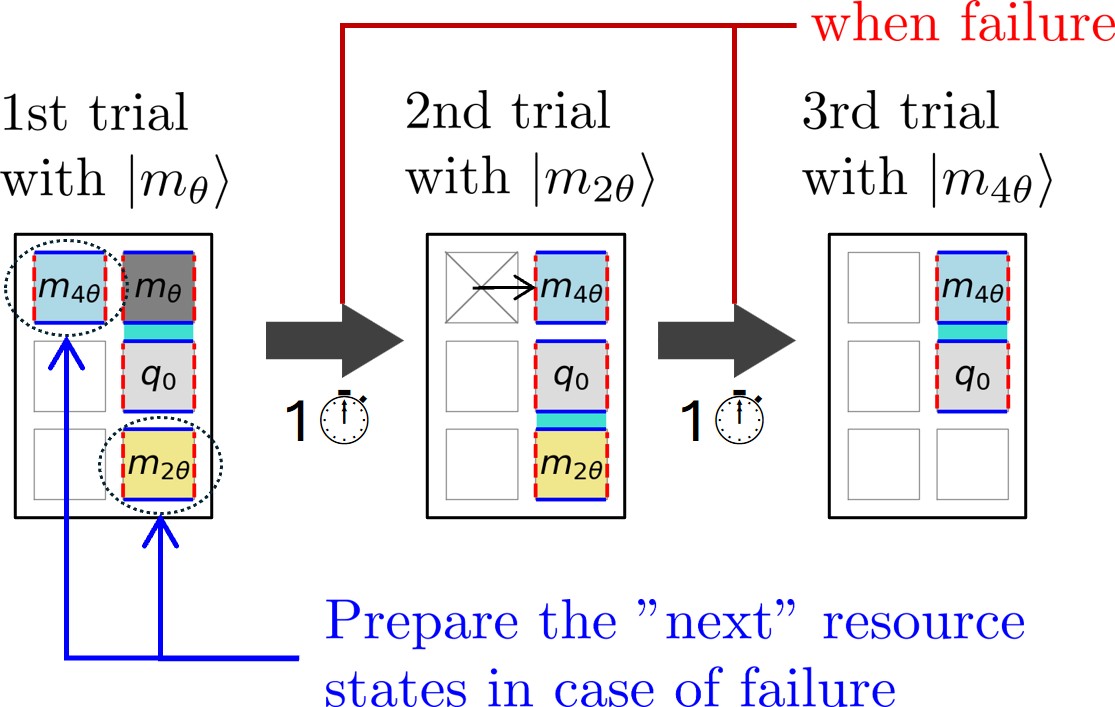}
\caption{\label{fig:qcoord}Schematic description of frequent attempts of analog rotations with an example of $\mathrm{Rz}(\theta)$ operation acting on qubit $q_0$. For instance, During the first trial of an analog rotation with $|m_\theta \rangle$ on the upper Z-edge of qubit $q_0$, we can prepare for the next rotation trial, by putting the resource state $ | m_{2\theta} \rangle$ on the lower Z-edge of $q_0$.
}
\end{figure}

\subsection{Qubit mapping}
Qubit mapping onto a surface code topology is one of the integral parts of overall quantum computation. It has been extensively studied in both NISQ computation with physical qubits and conventional surface code architecture with lattice surgery, to reduce spatial conflict of multiple operations. Currently, most works regarding surface code compilation is assuming that qubits are sparsely allocated (for instance, with one logical qubits per $2\times2$ spots). Such qubit topologies are good in terms of operation time efficiency, but not just efficient in terms of space efficiency. Given that we have quantum hardware with up to tens of thousands of qubits, where we call early-FTQC era, it should be better if we can put as many logical qubits as possible to explore the power of quantum computation which the classical computation is not able to realize. From this sense, we explore space-efficient type of qubit mapping. As we assume that one logical data qubit occupies just one patch space and no movement and edge reformation doesn't occur throughout overall computation, we set constraints for qubit mapping as follows:

\begin{enumerate}[(i)]
\item For each logical qubit, at least one X-edge and at least one Z-edge are accessible to unoccupied spots.
\item There is at least a single path from one of the X edges of logical qubit $q_1$ and one of the Z edges of logical qubit $q_2$ for $q_1,q_2 \in Q, q_1 \ne q_2$.
\end{enumerate}

\begin{figure}[htbp]
\centering
\includegraphics[scale=0.4]{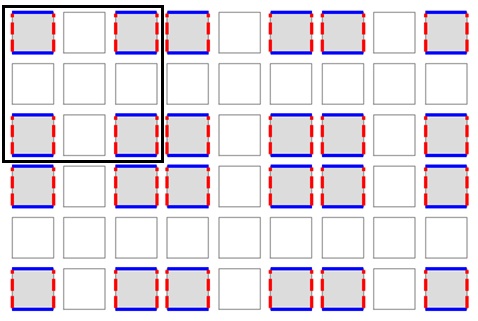}
\caption{\label{fig:densemap}An example of densest mapping of logical qubits. We can put $4mn$ logical qubits in $3m\times 3n$ spaces. $(m,n) = (3,2)$ in this figure.}
\end{figure}

The constraint (i) enables X(Z)-axis analog rotations for the qubits by executing X$\otimes$Z (Z$\otimes$Z) joint measurement with appropriate resource states without any movement and edge reformation of the logical data qubit. The constraint (ii) enables direct CNOT operation between the corresponding two logical qubits.
Our observation shows that the densest qubit mapping which meet these constraints is shown in \Cref{fig:densemap}. The ratio of the number of logical qubits to the number of spots are roughly $4:9$. Note that the maximal ratio can be realized when the patch spaces are composed of $3\times3$ blocks; otherwise, the maximum ratio is smaller.
In this paper, we randomly generate $\sim$200 random qubit topologies for each set of the number of qubits and logical qubit spaces. And we test the circuit execution performance in terms of the number of clock cycles needed for each random topologies. 


\section{\label{sec:res}Results}
We picked test circuits from QASMBench, which are transpiled with the gate set of \{
$\mathtt{CNOT}$, $\mathrm{Rx}(\pi/2)$, $\mathrm{Rz}(\theta)$\}. We further transpile these circuits with the gate set of \{$\mathtt{CNOT},\mathrm{Rx}(\phi),\mathrm{Rz}(\theta)$\}. Note that any kinds of single qubit gates are described with 
$$U_{\mathrm{single}} = \mathrm{Rz}(\theta_3)\mathrm{Rx}(\pi/2)\mathrm{Rz}(\theta_2)\mathrm{Rx}(\pi/2)\mathrm{Rz}(\theta_1)$$
and can be transpiled to

$$U_{\mathrm{single}} = \mathrm{Rz}(\theta_2)\mathrm{Rx}(\phi)\mathrm{Rz}(\theta_1).$$

\begin{figure}[htbp]
\centering
\includegraphics[scale=0.4]{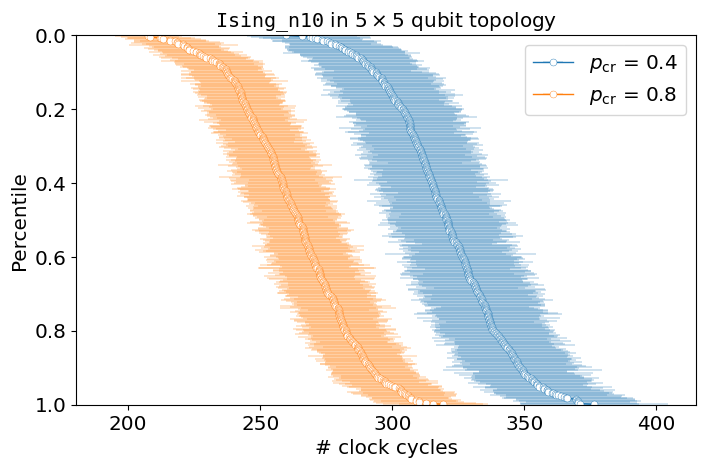}
\caption{\label{fig3}Resultant clock cycles ($T_\mathrm{clock}$) required to execute the circuit $\mathtt{Ising\_n10}$. We test the circuit simulation with 500 qubit topologies in $5\times5$ qubit space per setting of $p_{\mathrm{cr}} \in \{0.4,0.8\}$, which denotes success probability of resource state creation. A single data point corresponds to a single qubit topology and shows the average clock cycles among 50 operation trials. Shade denotes their standard deviations. The results are sorted in terms of avarage clock cycles, from fewest (with percentile 0.0) to greatest (with percentile 1.0)}
\end{figure}

As the resultant number of timesteps differs from trial to trial due to the probabilistic nature of analog rotations, we run the circuit simulation 50 times per set of \{qubit topology, target circuit\}. \Cref{fig3} shows the resultant numbers of clock cycles ($T_\mathrm{clock}$) to run the circuit $\mathtt{Ising\_n10}$ executed with 500 random qubit topologies generated in a qubit space of $ 5 \times 5 $. Note that all 50 operations are done successfully with our protocols. This shows that the best qubit topology results in $\sim$20\% fewer clock cycles for a single circuit run than the average one among all qubit topologies.

\begin{figure}[htbp]
\centering
\includegraphics[scale=0.36]{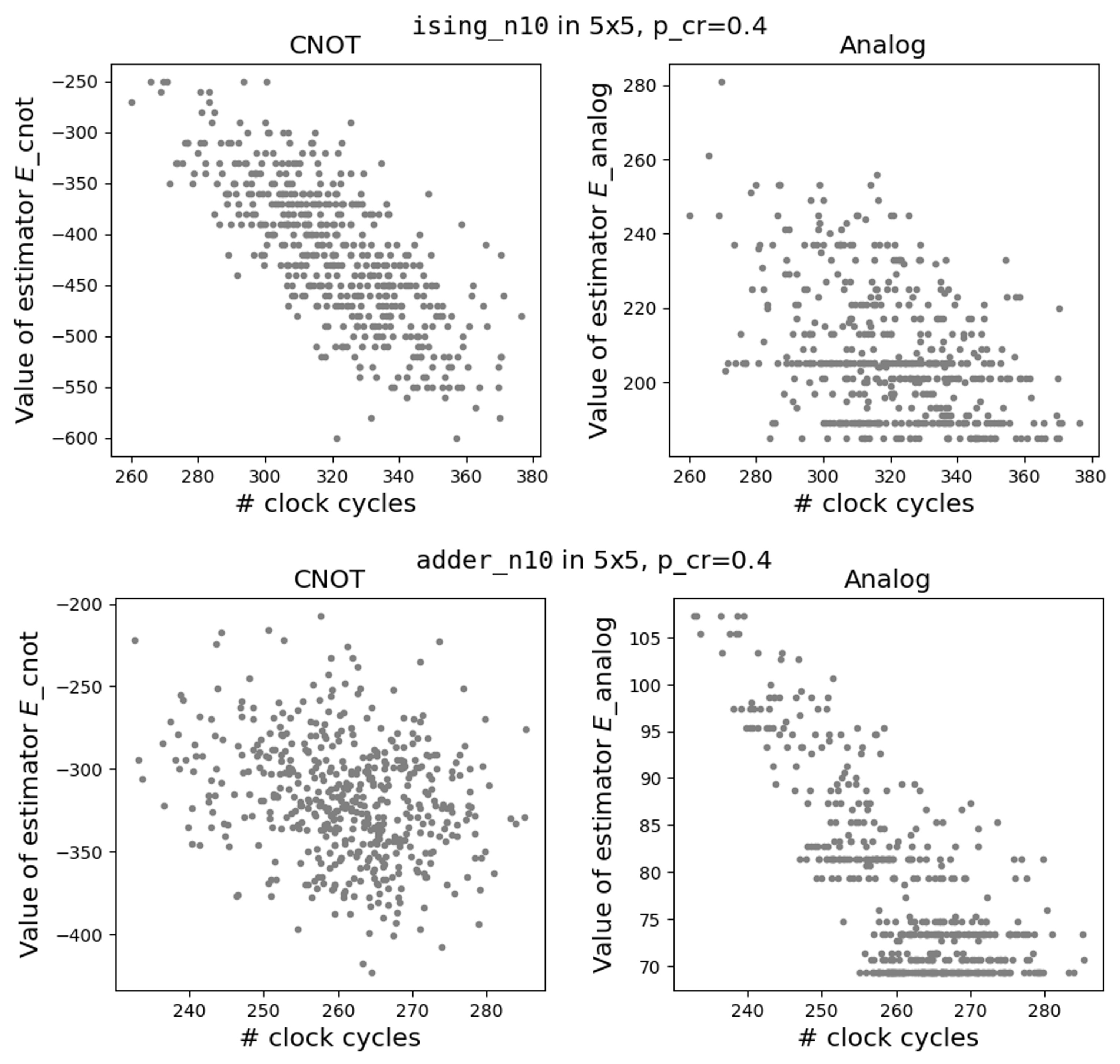}
\caption{\label{fig4}Dependence of average resultant clock cycles on each of the two performance estimators, $E_\mathrm{cnot}$ and $E_\mathrm{analog}$. A single point corresponds to experiment data for a logical qubit topology. Here the results of the circuit execitions of $\mathtt{ising\_n10}$ (upper figures) and $\mathtt{adder\_n10}$ (lower figures) with 500 random qubit topologies generated inside $5\times 5$ logical qubit space are shown.}
\end{figure}

We may want to choose the close-to-best qubit topology before actual operations. We can indeed choose the topology based on such simulation, but it will require a significant number of trials to determine a performance due to a large standard deviation of resultant clock cycles. To alleviate such computational workload, we introduce some performance estimator when a qubit topology and a target circuit are given. The key factors to affect the time performance are (a) the path length of CNOT operations and (b) how frequently we can try analog rotations (which includes how simultaneously we can try to create resource states, as well as how frequently we can do joint measurement between the logical data qubit and the resource state). We formulate the indicator based on each of the two factors:

\begin{align}
\label{eq:ind_edge}
E_\text{analog} = \sum_{q \in \{0, \cdots , n_q -1 \} } \biggl( \sum_{P \in  \{ \mathrm{X} , \mathrm{Z} \} }  e_{\text{open},q,P} \, n_{\text{ops},q,P} \biggr) 
\end{align}

\begin{align}
\label{eq:ind_path}
E_\text{cnot} = -1 \times \biggl( \sum_{\substack{c,t \in \{0, \cdots , n_q -1 \}\\ c\ne t}} n_{\text{cnot},c,t} \, L_{c,t} \biggr)
\end{align}

\noindent{where $e_{\text{open},q,P}$ denotes the number of open edges for qubit $q$ and Pauli $P$, $n_{\text{ops},q,P}$ the number of $P$-axis analog rotations acting on qubit $q$ in the target circuit, $n_{\text{cnot},c,t}$ the number of CNOT gates (over the whole circuit) between control qubit $c$ and target qubit $t$, and $L_{c,t}$ the path length from control qubit $c$ to target qubit $t$, respectively.}

Both of the estimators take features of both the qubit topology and the target circuit and evaluate affinity between them. 
The estimator $E_\mathrm{analog}$ in \Cref{eq:ind_edge} takes how many X/Z edges are accessible from free spots (from the qubit topology) and how many X/Z analog rotations are scheduled for each logical qubits (from the quantum circuit). It is based on the intuition that shorter path from the control qubit to the target qubit is preferred, when CNOT between them happens frequently. The estimator $E_\mathrm{cnot}$ in \Cref{eq:ind_path}, on the other hand, takes path length for each CNOT  (from the qubit topology) and the number of times of each CNOT (from the quantum circuit). It is based on the intuition that the more edges of the qubit resource states can access to, the more frequently we can try analog rotations. We plot the actual performance (i.e. the resultant avarage number of clock cycles) versus each of these performance estimators in \Cref{fig4}.

Regarding the examples shown in \Cref{fig4}, the performances' dependence on the two estimators shows a siginificant contrast: For the circuit $\mathtt{ising\_n10}$, for instance, the performance has a strong dependence on $E_\mathrm{cnot}$ with R-value of -0.737 and slightly weak dependence on $E_\mathrm{analog}$ with R-value of -0.405. For the circuit $\mathtt{adder\_n10}$, on the other hand, the performance has weak dependence on $E_\mathrm{cnot}$ with R-value of -0.245 and strong dependence on $E_\mathrm{analog}$ with R-value of -0.750. This difference may reflect some feature differences between the target circuits. 

To figure out such differences, we define the circuit density regarding the CNOT and single-qubit analog rotations. By defining a circuit layer which consists of CNOT layer "M" and single-qubit operation layer "S", the target circuit can be described as 
$${\mathrm{M}_{n_s} \mathrm{S}_{n_{s-1}} \mathrm{M}_{n_{s-1}} \mathrm{S}_{n_{s-2}} ... \mathrm{S}_1 \mathrm{M}_1 \mathrm{S}_0 \mathrm{M}_0},$$ 
where $n_s$ denotes the number of S layer. By limiting the number of times a single qubit is acted on by CNOT within one layer is up to 1, the maximum number of CNOTs in a single layer is $\lfloor q/2 \rfloor$. In addition, following the notion of Euler angles, we can fully describe any kind of single-qubit gates with up to 3 analog rotations. Therefore, we can say that the maximum numbers of single-qubit analog rotations and CNOT operations are $3q({n_s}+1)$ and $n_s \lfloor q/2 \rfloor$, respectively. From this value, we can define that CNOT density and single-qubit operation density as follows: 
\begin{align}
\label{eq:cnot_d}
D_\text{analog} = \frac{N_\text{analog}}{3q (n_s + 1)},
\end{align}
\begin{align}
\label{eq:cnot_d}
D_\text{cnot} = \frac{N_\text{cnot}}{n_s \lfloor q/2 \rfloor},
\end{align}

where $N_\text{cnot}$ and $N_\text{analog}$ denote the actual number of analog rotations in the target circuit and the actual number of CNOTs, respectively. Note that the density of $D_\text{analog}$ is correlated with $D_\text{cnot}$. More concretely, for the circuits with $N_\text{cnot}$ CNOTs, up to $N_\text{cnot}+1$ arbitrary single-qubit operations are sufficient to make the circuit maximally expressible. In such a sense, provided that any single-qubit gate can be described with up to three analog rotations, $N_{\mathrm{analog}} \le 3q+6 N_{\mathrm{cnot}}$, then 
\begin{align}
\label{eq:cnot_d}
D_{\mathrm{analog}} \le \frac{1}{n_s + 1} + \frac{2 D_\text{cnot} n_s \lfloor q/2 \rfloor}{(n_s + 1) q},
\end{align}

\begin{figure}[htbp]
\centering
\includegraphics[scale=0.35]{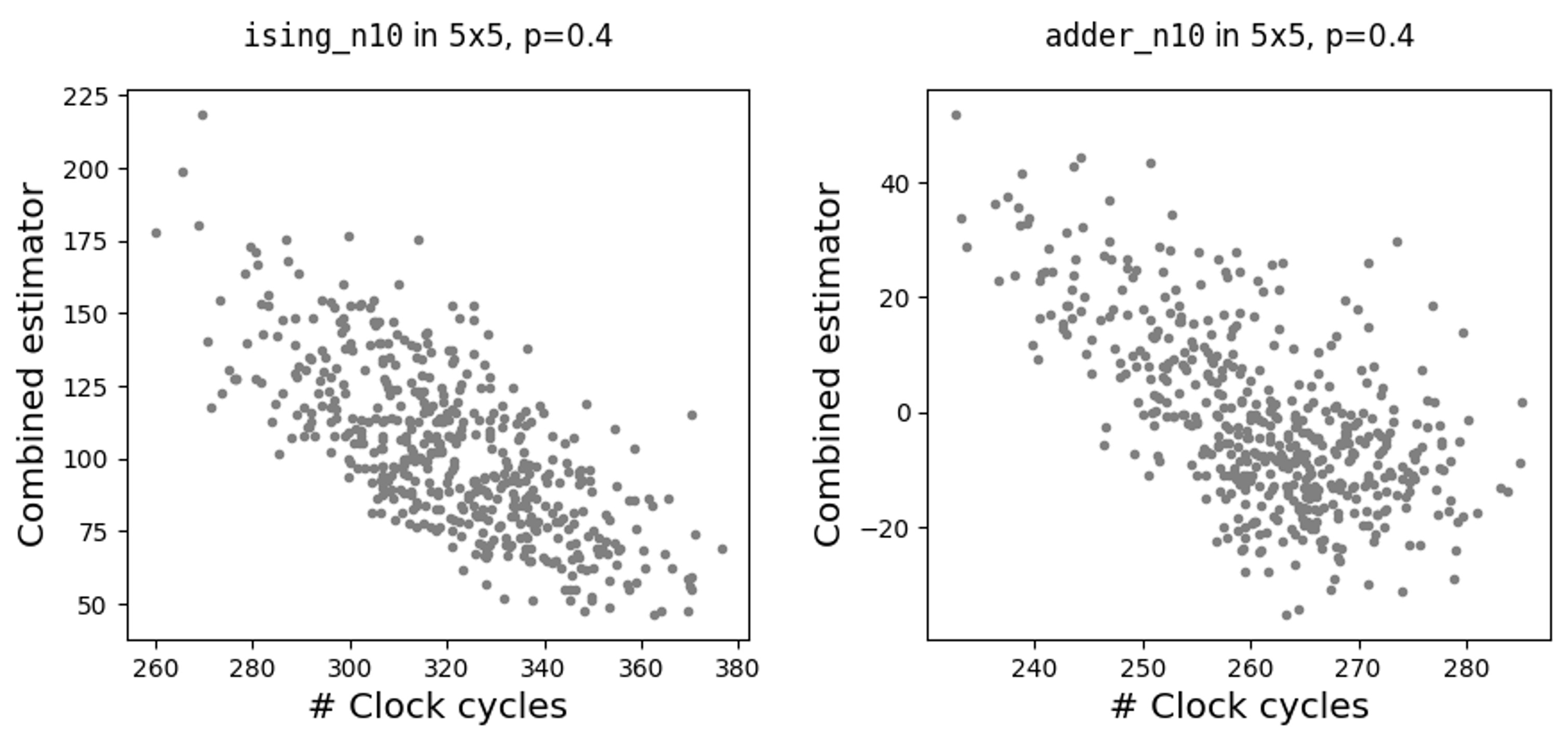}
\caption{\label{fig5}Dependence of average resultant clock cycles on the combined performance estimators, $E_\mathrm{comb}$ with $w=0.3$. A single point corresponds to experiment data for a logical qubit topology. Here the results of the circuit executions of $\mathtt{ising\_n10}$ (left figure) and $\mathtt{adder\_n10}$ (right figure) with 200 random qubit topologies generated inside $5\times 5$ logical qubit space are shown.}
\end{figure}

 \Cref{tab4} summarizes the resultant R-value for all circuits we tested, as well as the circuit densities. We can see some trade-offs between the correlation strength between $T_\mathrm{clock}$ and $E_\mathrm{analog}$ and that between $T_\mathrm{clock}$ and $E_\mathrm{cnot}$. For example, circuits with greater $D_\mathrm{analog}$ and $D_\mathrm{cnot}$ demonstrate good correlation strengths between $T_\mathrm{clock}$ and $E_\mathrm{cnot}$. This implies that the lengths of CNOT paths on the logical qubit space affect the performance. The length of CNOT path is the number of spots to use for the CNOT, therefore because this length value equals to the number of free spots to use in the CNOT routings. This can limit the space of resource state creations, and such conditions become more severe when $D_\mathrm{cnot}$ is large. In this sense, the correlation between $T_\mathrm{clock}$ and $D_\mathrm{analog}$ is likely to be weak. By contrast, circuits with lower $D_\mathrm{analog}$ and $D_\mathrm{cnot}$ demonstrate good correlation strengths between the number of clock cycles and $E_\mathrm{analog}$. For such circuits, the number of free spots occupied by CNOT routing is fewer, so that we can reserve more free spots for resource state creation. In this sense, the correlation between the number of clock cycles and $D_\mathrm{analog}$ is likely to be stronger than between the number of clock cycles and $D_\mathrm{cnot}$.
 Further, we may want a single estimator which can estimate the performance regardless of circuit features. To do this, we introduce a combined estimator:
 \begin{align}
\label{eq:ind_comb}
E_\text{comb} = E_\text{analog} + w E_\text{cnot} ,
\end{align}
where $w\in \mathbb{R}$ denotes a weight. As can be seen in \Cref{fig5}, the performance estimator $E_\mathrm{comb}$ have good correlation with actual performance for both $\mathtt{ising\_n10}$ and $\mathtt{adder\_n10}$. \Cref{tab4} also shows the R-values between $T_\mathrm{clock}$ and $E_\text{comb}$ with $w=0.3$. It can be seen that all absolute values of the R-values are above 0.5, suggesting that $T_\mathrm{clock}$ and $E_\text{comb}$ have good correlations.
 This observation is a key to find an appropriate logical qubit topology without the actual circuit-run simulation. 

\begin{figure}[htbp]
\centering
\includegraphics[scale=0.16]{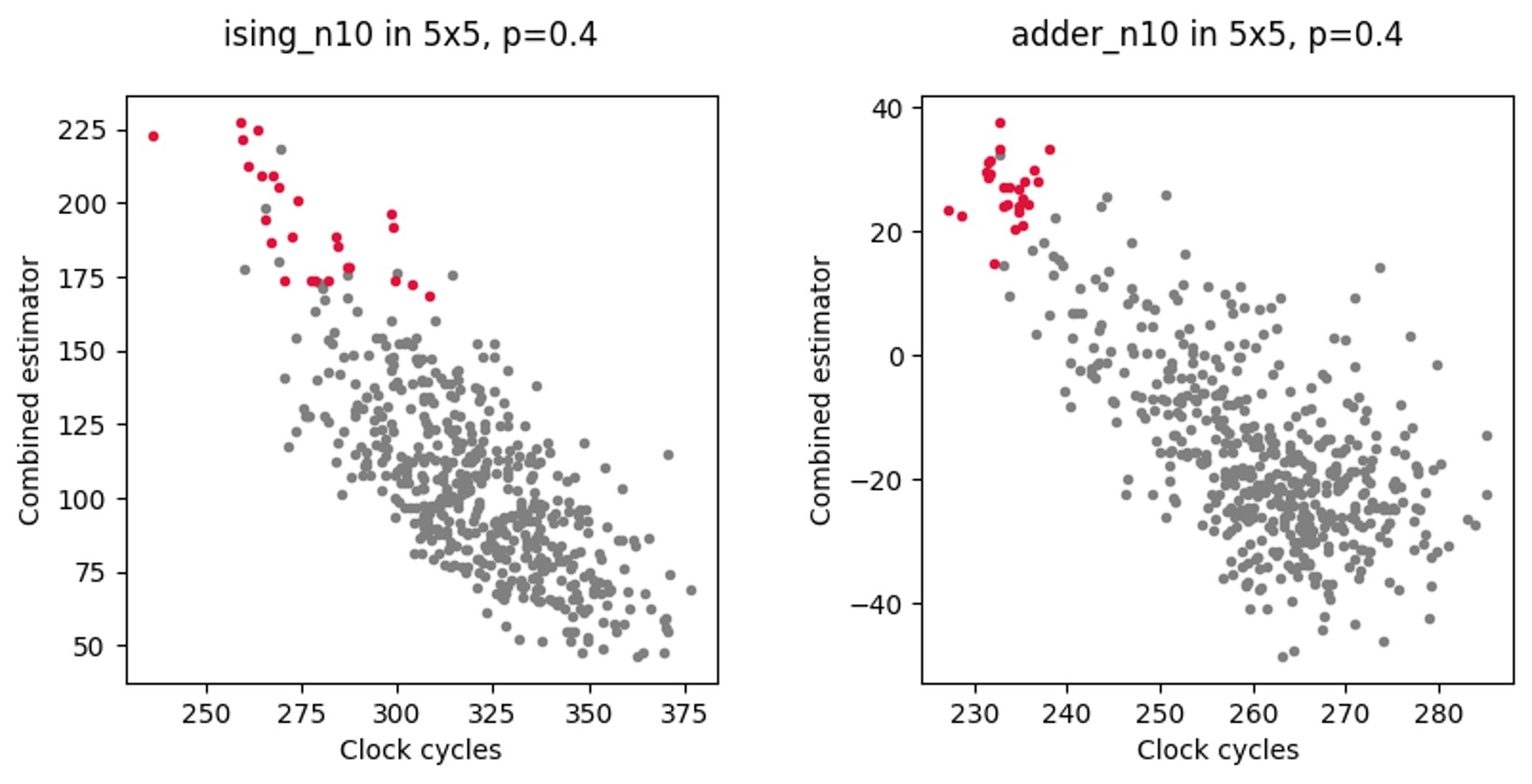}
\caption{\label{fig6}Dependence of average resultant clock cycles on the combined performance estimators, $E_\mathrm{comb}$ with $w=0.3$. A single point corresponds to experimental data with a single logical qubit topology. The red points denote the results of the circuit executions of $\mathtt{ising\_n10}$ (left figure) and $\mathtt{adder\_n10}$ (right figure) with qubit topologies which have top-25 highest values of $E_\mathrm{comb}$ among 5,000 randomly-generated topologies inside $5\times 5$ logical qubit space. The gray points denote that the results with 500 randomly-generated qubit topology (the same data as those shown in \Cref{fig5})}
\end{figure}

\nopagebreak
\begin{table*}[htbp]
\centering
\caption{\label{tab4} The resultant R-values and $T_\text{clock}$ for each target circuit and logical qubit space.} 
\begin{tabular}{|wc{1.6cm}|wc{1.0cm}|wc{1.0cm}|wc{1.5cm}|wc{0.6cm}|wc{1.2cm}|wc{1.2cm}|wc{1.2cm}|wc{1.0cm}|wc{1.0cm}|}
\hline
\multirow{2}{*}{\begin{tabular}{c} Circuit \\name \end{tabular}} &\multirow{2}{*}{$D_\mathrm{analog}$} & \multirow{2}{*}{$D_\mathrm{cnot}$}& \multirow{2}{*}{\begin{tabular}{c} Logical \\qubit space \end{tabular}}  & \multirow{2}{*}{$p_\mathrm{cr}$} &
\multicolumn{3}{|c|}{R-values} & 
\multicolumn{2}{|c|}{$T_{\text{clock}}$}
\\
\cline{6-10}
  &    &&& &
$E_\text{analog}$ & 
$E_\text{cnot}$ & 
$E_\text{comb}$ & 
Avg. & 
Min.
\\
\hline \hline
\multirow{4}{*}{$\mathtt{dnn\_n8}$} &\multirow{4}{*}{0.3628} &\multirow{4}{*}{1.0} & \multirow{2}{*}{$5\times5$} & 0.4 & -0.3758 & -0.7550 & -0.6827 & 586.8 & 485.9  \\ \cline{5-10}
                                && &                                & 0.8 & -0.3948 & -0.7875 & -0.7135 & 504.0 & 408.3 \\ \cline{4-10}
                                 &&& \multirow{2}{*}{$6\times6$}    & 0.4 & -0.4092 & -0.5867 & -0.5686 & 486.2 & 402.4 \\ \cline{5-10}
                                 &&&                                & 0.8 & -0.4272 & -0.6235 & -0.5992 & 418.2 & 345.8  \\ \hline \hline
                                
\multirow{4}{*}{$\mathtt{qpe\_n9}$} &\multirow{4}{*}{0.0614} &\multirow{4}{*}{0.2275} & \multirow{2}{*}{$5\times5$} & 0.4 & -0.8457 & -0.1994 & -0.6271 & 233.9 & 207.0  \\ \cline{5-10}
                                 &&&                                & 0.8 & -0.8436 & -0.2630 & -0.6689 & 201.0 & 177.9 \\ \cline{4-10}
                                 &&& \multirow{2}{*}{$6\times6$}    & 0.4 & -0.9214 & -0.2728 & -0.7610 & 216.2 & 190.3 \\ \cline{5-10}
                                 &&&                                & 0.8 & -0.9435 & -0.2695 & -0.7730 & 188.0 & 166.5  \\ \hline \hline

\multirow{4}{*}{$\mathtt{ising\_n10}$} &\multirow{4}{*}{0.3042} &\multirow{4}{*}{0.9000} & \multirow{2}{*}{$5\times5$} & 0.4 & -0.4055 & -0.7372 & -0.7229 & 320.9 & 259.9  \\ \cline{5-10}
                                 &&&                                & 0.8 & -0.3882 & -0.7810 & -0.7426 & 264.0 & 207.5 \\ \cline{4-10}
                                 &&& \multirow{2}{*}{$6\times6$}    & 0.4 & -0.4084 & -0.6282 & -0.6643 & 255.0 & 200.5 \\ \cline{5-10}
                                 &&&                                & 0.8 & -0.3834 & -0.6820 & -0.6857 & 210.6 & 164.2  \\ \hline \hline

\multirow{4}{*}{$\mathtt{adder\_n10}$}& \multirow{4}{*}{0.0413} &\multirow{4}{*}{0.2364} & \multirow{2}{*}{$5\times5$} & 0.4 & -0.7551 & -0.2454  & -0.5935 & 261.1 & 232.7  \\ \cline{5-10}
                                 &&&                                & 0.8 & -0.7255  & -0.2412 & -0.5737 & 239.1 & 209.7 \\ \cline{4-10}
                                 &&& \multirow{2}{*}{$6\times6$}    & 0.4 & -0.8758 & -0.2851 & -0.6758 & 242.6 & 215.7 \\ \cline{5-10}
                                 &&&                                & 0.8 & -0.8668 & -0.2750 & -0.6637 & 223.5 & 197.9  \\ \hline \hline

\multirow{4}{*}{$\mathtt{dnn\_n16}$} &\multirow{4}{*}{0.3628} &\multirow{4}{*}{1.0} & \multirow{2}{*}{$6\times6$} & 0.4 &  0   & -0.9416 & -0.9416 & 795.2 & 722.7  \\ \cline{5-10}
                                 &&&                                & 0.8 & 0     & -0.9313 & -0.9313 & 742.0 & 611.8 \\ \cline{4-10}
                                 &&& \multirow{2}{*}{$7\times7$}    & 0.4 & -0.1823 & -0.7814  & -0.7043 & 708.1 & 590.6 \\ \cline{5-10}
                                 &&&                                & 0.8 & -0.1372 & -0.7043  & -0.6697 & 633.4 & 513.1  \\ \hline \hline

\multirow{2}{*}{$\mathtt{qft\_n18}$} &\multirow{2}{*}{0.1318} &\multirow{2}{*}{0.5152} & \multirow{2}{*}{$7\times7$} & 0.4 &  -0.4205   & -0.3903 & -0.5237 & 737.9 & 654.4  \\ \cline{5-10}
                                 &&&                                & 0.8 & -0.3859    & -0.4276 & -0.5353 & 637.3 & 556.7 \\ \hline \hline

\multirow{2}{*}{$\mathtt{qft\_n29}$} &\multirow{2}{*}{0.1291} &\multirow{2}{*}{0.5091} & \multirow{2}{*}{$9\times9$} & 0.4 &  -0.0619   & -0.5526 & -0.5321 & 1478.9 & 1286.1  \\ \cline{5-10}
                                 &&&                                & 0.8 & -0.0530     & -0.5384 & -0.5158 & 1320.5 & 1107.4 \\ \hline \hline

\multirow{2}{*}{$\mathtt{qugan\_n39}$} &\multirow{2}{*}{0.0319} &\multirow{2}{*}{0.1488} & \multirow{2}{*}{$10\times10$} & 0.4 &  -0.1271   & -0.5775 & -0.5890 & 692.8 & 627.7  \\ \cline{5-10}
                                 &&&                                & 0.8 & -0.1010    & -0.5895 & -0.5960 & 622.5 & 563.0 \\ \hline \hline

\multirow{2}{*}{$\mathtt{adder\_n64}$}& \multirow{2}{*}{0.0139} &\multirow{2}{*}{0.0786} & \multirow{2}{*}{$13\times13$} & 0.4 &  -0.4458   & -0.4939 & -0.5345 & 1003.2 & 898.5  \\ \cline{5-10}
                                 &&&                                & 0.8 & -0.4039    & -0.5051 & -0.5406 & 931.6 & 845.6 \\ \hline \hline

\multirow{2}{*}{$\mathtt{ising\_n66}$} &\multirow{2}{*}{0.2414} &\multirow{2}{*}{0.9848} & \multirow{2}{*}{$13\times13$} & 0.4 &  -0.0828   & -0.6072 & -0.6100 & 160.0 & 120.7  \\ \cline{5-10}
                                 &&&                                & 0.8 & -0.0675     & -0.5443 & -0.5459 & 149.9 & 110.7 \\ \hline

\end{tabular}
\end{table*}

Then, we tried to generate a logical qubit topologies which has greater performance estimators. For given quantum circuits from QASMBench and patch spaces, we generate 5,000 random topologies and pick 25 topologies among them which show greater performance estimators $E_\text{comb}$ than the others do. By testing the actual performance $T_\mathrm{clock}$ as shown in \Cref{fig6}, the picked 25 qubit topologies demonstrate fewer $T_\text{clock}$ to run the circuit than fully random qubit topologies do.

\section{\label{sec:conclusion}Conclusion}
We propose the circuit operation protocol with logical qubits with surface code, lattice surgery and STAR architecture. This protocol is applicable for general kinds of circuits.
\begin{itemize}
  \item{To run the circuit without failure, we set a prioritizing rule of element operation with lattice surgery. CNOT operations are the most prioritized among all the surface-code operations.}
  \item{We introduce protocols to reduce the clock-cycle overhead induced by probabilistic procedures of resource state creations and consumptions. To reduce the time overhead of resource-state creations, we incorporate a method to try resource state creations at multiple spots in parallel. In doing this, we also incorporate a protocol to maximize the number of spots for resource state creation via optimization with QUBO.}
  \item{To reduce the clock-cycle overhead for analog rotations, we incorporate an protocol to reduce the intervals between the rotation trials.}
  \item{To map the logical qubits into the given space in an efficient way in terms of clock cycles, we introduce two kinds of indicators which can predict the performance without actual simulations of circuit runs.}
\end{itemize}
We leave the remaining problem as a future work:
\begin{itemize}
  \item{While we let the logical data qubit be stationary in the qubit space in this paper, the efficiency of circuit run can be further improved when we allow logical qubit movement and reformation. More concretely, when we devide the circuit into some segments each of which consists of a few cycles, it can be the case that appropriate qubit mapping is different between them. In such cases, we can insert protocols of changing the qubit topology between adjecent segments, at the cost of clock cycles which is are needed for logical qubit movement.}
  \item{There are another major type of circuit representation, proposed by Litinski et al \cite{Litinski_2019_Game}. With this method, a circuit can be represented by cycles with potentially global arbitrary-angle rotation followed by global Clifford measurements. While a single analog rotation comes to act on more logical qubits, our observation shows that we can still do simultaneous operations. However, we have to set a path between the qubits to be acted on, which will be occupied until the target analog rotation is successful. This can likely block other global operations especially when the logical qubits are densely mapped. In this case, the choice of qubit mapping will be more important, and likely we will need some qubit-topology changes between circuit segments.}
\end{itemize}

\addtolength{\textheight}{-16.3cm}   




\section*{ACKNOWLEDGMENT}

We thank Yutaro Akahoshi and Jun Fujisaki for fruitful discussions.


\bibliographystyle{unsrt}
\bibliography{bib}

\end{document}